# É tempo de dizer **eureka!** no ensino de Física

José M. L. Figueiredo

**Resumo:** No espírito do Ano Internacional da Física, o autor propõe a substituição dos métodos de ensino baseados na exposição verbal e na execução de instruções previamente fornecidas por processos que envolvam activamente o estudante.

## I. Introdução

É frequente cada geração inquietar-se com o facto, aparente ou não, dos padrões educativos estarem a baixar.[1] No entanto, a sociedade acaba por afastar muitos dos seus jovens, em particular durante a adolescência, da procura de conhecimento, em especial do científico. Em resultado a influência da pseudociência na vida dos cidadãos, mesmo nos mais escolarizados, não esmorece:

"Não só nos lares camponeses, mas também nos arranha-céus das cidades, coabitam o século XX e o século XIII. Cem milhões de pessoas utilizam a electricidade e ainda acreditam nos poderes mágicos dos signos e nos exorcismos ... As estrelas de cinema vão aos médiuns. Os aviadores que pilotam máquinas maravilhosas criadas pelo génio do homem usam amuletos nas camisolas. Que reservas inesgotáveis possuem eles de trevas e de ignorância".[2]

Mesmo indivíduos com formação científico-tecnológica atribuem crédito às teorias da pseudociência. Porque será?

## II. "Não basta dar o peixe, é preciso ensinar a pescar"

Com raras excepções, os métodos de ensino dominantes na ciência são idênticos aos usados pelas diferentes variantes da pseudociência, limitando-se a relatar as descobertas e os produtos da ciência, desprezando a observação e a investigação crítica metódica dos fenómenos e conceitos. Como pode assim o estudante distinguir a Ciência da falsa ciência?

O método de ensino expositivo tem perpetuado o ponto de vista segundo o qual "ensinar bem" consiste quase inteiramente em narrações verbais claras e na repetição de procedimentos demonstrativos dos princípios abordados nas aulas "teóricas". Estudos sobre o ensino de Física mostram claramente que grande parte dos alunos dos ensinos secundário e universitário não interiorizam correctamente os conceitos físicos e são incapazes de relacionar entre si muitos dos tópicos abordados. Como explicar que os estudantes após terem frequentado um curso introdutório de Física de nível universitário com uma elevada componente de Mecânica fiquem surpreendidos quando, por exemplo, se lhes revela que a aceleração da gravidade a 500 km da superfície terrestre é cerca de oitenta e seis porcento do valor característico à superfície; que a acção gravitacional do planeta Marte, por exemplo, na altura do nosso nascimento é da mesma ordem de grandeza da do obstetra; que a situação de "imponderabilidade" vivida nas estações espaciais, quase sempre interpretada como sendo consequência do grande afastamento da Terra ("microgravidade"), se deve ao facto de tanto a estação orbital como os corpos no seu interior estarem em queda livre? Quantos alunos interrelacionam temas como a convecção, a gravitação e a queda livre quando confrontados perante a diferença de forma da chama de uma vela na Terra e na situação de "imponderabilidade", ou recorrem ao princípio acção-reacção para justificar a dificuldade em caminhar numa nave espacial em órbita? O mesmo acontece com os outros temas leccionados. Por exemplo, quantos alunos conhecem a razão de o vidro da porta do forno microondas conter uma grelha metálica?

[1] Um dos mais antigos ensaios na história da humanidade, da Suméria, datado de há mais de quatro mil anos, lamenta o facto de os mais jovens serem mais ignorantes do que a geração precedente.

[2] Atribuído a Leon Trotski (pouco antes de Hitler chegar ao poder).

Outro aspecto associado ao anterior, mas não menos relevante, é a quase incapacidade dos estudantes fazerem estimativas quantitativas, ainda que grosseiras, de situações correntes. Poucos alunos têm noções de ordem de grandeza, experiência em comparação de tamanhos, de escalas e de dimensões. Por exemplo, quantos estudantes são capazes, usando apenas papel e lápis, de determinar o número típico de carros envolvidos numa fila de trânsito com dois quilómetros numa via de três faixas? Quantos litros de gasolina são necessários para satisfazer as necessidades anuais do parque automóvel português?

A capacidade de analisar criticamente fenómenos ou questões físicas, de desenvolver modelos e de compreensão conceptual requerem a interiorização de uma variedade de princípios físicos e uma panóplia de habilidades que raramente são introduzidos até ao curso de pós-graduação ou na actividade profissional.

"Na Universidade, os meus sonhos tornaram-se realidade: encontrei professores que não só compreendiam a ciência, mas também eram capazes de a explicar ... Foi com Kuiper que pela primeira vez me entusiasmei com aquilo a que se chama um 'cálculo nas costas do envelope': ocorre-nos uma explicação possível para um problema, tira-se da gaveta um envelope velho, faz-se apelo aos nossos conhecimentos de física fundamental, escrevinham-se umas quantas equações aproximadas no envelope, substituem-se letras por valores numéricos prováveis e vê-se se a resposta consegue explicar o problema. Se assim não for, procura-se uma explicação diferente. Elimina-se o absurdo com a maior das facilidades".[3]

### III. A MERA ACUMULAÇÃO DE FACTOS OU PROCEDIMENTOS NÃO CONSTITUI UMA CIÊNCIA

As causas da debilidade de análise e de interrelação demonstradas pelos estudantes tem várias origens. Alguns estudos atribuem estas incapacidades, em parte, ao facto de muitos dos tópicos abordados serem tratados como assuntos independentes, ignorando ou mesmo evitando, muitas vezes, os exemplos que permitiriam relacionar os diferentes temas. Muitos programas dos cursos introdutórios de Física resumem-se, quase exclusivamente, ao tratamento minucioso da Mecânica Clássica, da Termodinâmica, e do Electromagnetismo, muitas vezes sem grande relevância quotidiana. Temas como a Óptica, a Acústica, a Relatividade, a Física Quântica e a Estrutura da Matéria muito raramente são abordados. E, em geral, os temas são apresentados a um nível teórico demasiado fechado. Para além disso, no ensino superior as metodologias empregues exigem quase sempre uma preparação matemática para além da álgebra e da trigonometria, e no ensino secundário a análise quantitativa dos assuntos nem sempre é devidamente estimulada.

Na sala de aula raramente se promove o espírito crítico - tendência para não admitir nenhuma asserção sem reconhecer a sua legitimidade ou fundamento - como a atitude intelectual de excelência e característica basilar da ciência. E embora os exemplos e os problemas incluídos na actual geração de textos introdutórios aparentem ter sido modernizados, na maior parte dos casos, mesmo os de aspecto moderno, são simplesmente reformulações de problemas clássicos retirados de textos mais antigos. A artificialidade dos exemplos e dos problemas e a introdução muito limitada ao extenso conjunto de perícias profissionais associadas à Física fornecem ao estudante uma imagem enganadora da Física e é muitas vezes desmotivante para o aluno médio. Mesmo os estudantes mais brilhantes, capazes de resolver problemas clássicos com facilidade, demonstram, frequentemente, falhas de compreensão conceptual e dificuldades em analisar fenómenos físicos ou questões científicas empregando o "método" científico, e fazem pouca ou nenhuma ideia sobre os modo de proceder dos físicos como por exemplo no desenvolvimento de modelos. O "método" científico não é propriamente um método, é acima de tudo um conjunto de atitudes alicerçadas no primado da observação, da razão e da experimentação, cujo objectivo é aumentar o grau de conhecimento sobre a forma como a Natureza funciona.

Acredito firmemente na importância da experimentação laboratorial e conceptual como parte integrante da educação de um estudante, e partilho a convicção que as actividades lectivas devem ter como propósito fundamental a disseminação da atitude científica e promover o desempenho experimental e intelectual do estudante, tendo sempre em consideração os aspectos práticos da área em que se insere a sua especialização. Actualmente as actividades laboratoriais consistem, maioritariamente, na simples verificação de alguns dos princípios tratados nas aulas "teóricas", em que o bom desempenho apenas requer que se siga um conjunto de procedimentos que leva ao resultado final - não há descoberta e, portanto, não há, em geral, motivação! É essencial que as aulas laboratoriais permitam ao estudante descobrir relações entre grandezas, desenvolver modelos qualitativos e/ou matemáticos a partir de observações, para além de assimilar as técnicas e as atitudes laboratoriais adequadas.

O domínio do "método" científico, por muito maçador e hostil que possa parecer, é mais importante que o relato das descobertas científicas ou dos conhecimentos científicos. É fundamental ensinar os estudantes a elaborar modelos do mundo físico que nos rodeia. Este treino permitirá ao aluno compreender mais facilmente como as teorias leccionados foram sendo criadas. A aprendizagem de conceitos através da realização de experiências, laboratoriais ou conceptuais, e o desenvolvimento de modelos que podem ser aplicados a muitas situações no futuro, noutros tópicos ou na actividade profissional, permitirá ao estudante contribuir com novas ideias e soluções e enfrentar situações inesperadas sem misticismo e com menor receio.

---

[3] Carl Sagan, *Um Mundo Infestado de Demónios*, Gradiva, 1997. Gerard Kuiper, astrónomo americano, 1905-1973, considerado o pai da ciência planetária moderna.

## IV. A FÍSICA É IMPORTANTE PARA TODOS E DEVE SER INTERESSANTE PARA MUITOS MAIS

A grande maioria dos alunos dos ensinos politécnico e universitário que procurem carreiras nas áreas da engenharia, da informática, da medicina e da tecnologia médica, da microbiologia, da química e bioquímica, ou de outras opções profissionais na área das tecnologias, frequentam cursos cujo currículo compreende apenas uma cadeira introdutória de Física, em geral, 2 semestres no 1º ano de licenciatura. (Muitos dos que clamam que a formação básica em Ciências é essencial não se coíbem de propor cursos introdutórios de Física com apenas um semestre.[4]) A preocupação fundamental destes alunos, no que respeita à Física, relaciona-se com o entendimento do papel que esta desempenha na compreensão das áreas de conhecimento e das tecnológicas em que a sua formação se insere.

Embora os caloiros possam não estar conscientes disso, aos diplomados – elite de potenciais inovadores, empreendedores e dirigentes – a sociedade exige, cada vez mais, que executem actividades ou desempenhem funções em áreas não directamente relacionadas com a sua formação, pois será pouco provável que permaneçam no mesmo tipo de tarefa durante períodos de tempo consideráveis, podendo ter que assumir repentinamente funções de responsabilidade em áreas de trabalho diversas. Acresce ainda que a convivência com problemas complexos que exigem solução rápida tende a ser elevada, pelo menos no contexto de organizações que aspiram a liderar internacionalmente um dado ramo de actividade.

É pois essencial que o diplomado seja capaz de ler e de assimilar literatura em outros tópicos, para além do da especialidade. As capacidades de avaliar, interpretar e compreender efectivamente conceitos básicos de Física e a importância da ciência e da tecnologia facilitarão certamente o desempenho dos diplomados na estrutura produtiva, possibilitando uma maior transformação dos conhecimentos científicos e tecnológicos em benefícios práticos para a economia e sociedade portuguesas.

## V. CONCLUSÃO

O objectivo final de qualquer sistema de ensino plural, para além da formação especializada, é contribuir efectivamente para o desenvolvimento do espírito crítico, inventivo e criador do estudante, estimulando o seu interesse pela Cultura e pela Ciência. Este propósito exige o abandono do método de ensino expositivo e repetitivo, por forma a diminuir a diferença entre o que se pretende ensinar e o que é realmente aprendido, aumentando a literacia científica do cidadão e, em particular, diminuindo o iletrismo científico das elites.[5] Independentemente da área de conhecimento e do grau de ensino, é fundamental que a escola estimule simultaneamente o discernimento e a capacidade de utilizar e coordenar a informação recebida. É indispensável pôr os estudantes a pensar, tendo sempre presente que "ensinar é apenas ajudar a aprender",[6] e introduzir a **descoberta** no ensino de Física.

---

[4] Ao mesmo tempo que contestam a decisão da tutela quando promove a subalternização das ciências experimentais nos currículos do ensino secundário.

[5] "*O maior mal não é o analfabetismo, é o iletrismo das classes dirigentes*," Ricardo Jorge, médico e humanista, 1858-1939.

[6] E. Mazur, "*Ensinar é apenas ajudar a aprender*," Gazeta de Física **26**, Fasc. 1, 2003, p. 18-22.